\documentclass[a4paper]{jpconf}
\usepackage{graphicx}
\def\specchar#1{{\sc #1}}

\def\CaIIH{\mbox{Ca\,\specchar{ii}\,\,H}}       

\usepackage[square,sort&compress]{natbib}
\usepackage{color}

\begin{document}
\title{Magnetohydrodynamic waves driven by $p$-modes}

\author{Elena Khomenko and Irantzu Calvo Santamaria}

\address{Instituto de Astrof\'{\i}sica de Canarias, 38205 La Laguna, Tenerife, Spain}
\address{Departamento de Astrof\'{\i}sica, Universidad de La Laguna, 38205, La Laguna, Tenerife, Spain}

\ead{khomenko@iac.es, irantzu@iac.es}

\begin{abstract}
Waves are observed at all layers of the solar atmosphere and the magnetic
field plays a key role in their propagation. While deep down in the
atmosphere the $p$-modes are almost entirely of acoustic nature, in the upper
layers magnetic forces are dominating, leading to a large variety of new wave
modes. Significant advances have been made recently in our understanding of
the physics of waves interaction with magnetic structures,  with the help of
analytical theories, numerical simulations, as well as high-resolution
observations. In this contribution, we review recent observational findings
and current theoretical ideas in the field, with an emphasis on the following
questions: (i) Peculiarities of the observed wave propagation in network,
plage and facular regions; (ii) Role of the mode transformation and
observational evidences of this process; (iii) Coupling of the photosphere,
chromosphere, and above by means of waves propagating in magnetic structures.
\end{abstract}

\vspace{-1cm}

\section{Introduction}
\vspace{0.3cm}

Solar $p$-modes are generated by turbulent convection just beneath the
photosphere and propagate through the solar interior \cite{Lighthill1952,
Stein1967, Goldreich+Kumar1990, Goldreich+Murray+Kumar1994, Bi+Li1998,
Musielak+etal1994, Nordlund+Stein2001, Stein+Nordlund2001}. They also leak
power to the atmospheric layers $-$ photosphere, chromosphere and corona $-$
and interact with magnetic structures present in these layers. Since the
force balance changes with height from a gas pressure dominated region below
the photosphere into a magnetically dominated region in the corona, the
nature of waves changes as well and mode transformation occurs. These
$p$-modes leaking to the upper atmosphere are a possible source of
chromospheric/coronal heating. Besides, measuring and modeling the wave
properties provides us with sensible indicators of the thermodynamic and
magnetic structure of the atmosphere where they propagate \cite{Gizon2005,
Chaplin2008, Basu2008, Moradi2010}. This article reviews the current advances
in the field of atmospheric waves and oscillations driven by solar $p$-modes
with an emphasis on their interaction with local magnetic fields.

Observationally, sunspot waves have been a usual and relatively easy target,
due to the clear influence of the strong sunspot's magnetic field on them.
Photospheric waves in the umbra and penumbra have been detected over decades,
as well as their propagation to the chromosphere in the form of umbral
flashes (see e.g., \cite{Lites1984, Lites1986, Lites1988,
Lites+Thomas+Bogdan+Cally1998, Abdelatif+Lites+Thomas1986, Maltby+etal1999,
Maltby+etal2001, Brynildsen+etal2000, Brynildsen+etal2002,
Christopoulou+etal2000, Christopoulou+etal2001, Rouppe+etal2003,
Centeno+etal2006a, Tziotziou+etal2006, Tziotziou+etal2007,
LopezAriste+etal2001, Felipe+etal2010b}, and many others). Sunspot magnetic
fields are easily detectable via polarimetric measurements by many
instruments \cite{Borrero+Ichimoto2011}, and the measured parameters of
oscillations and of the magnetic field can be linked. The general aspects of
physics of these waves {(such as the different types of waves, their physical
properties, the conditions requires for transforming one into the other,
etc.) are relatively well understood nowadays (for reviews on sunspot waves
before 2008, see \cite{Bogdan+Judge2006, Khomenko2009}). Numerical modeling
of sunspot waves has helped to explain many details
\cite{Cally+Bogdan+Zweibel1994, Cally+Bogdan1997, Rosenthal+etal2002,
Rosenthal+Julien2000, Bogdan+etal2003, Khomenko+Collados2006,
Khomenko+Collados2008, Khomenko+Collados2009, Khomenko+etal2009,
Parchevsky+Kosovichev2007, Parchevsky+Kosovichev2007b,
Parchevsky+Kosovichev2009, Hanasoge2008, Schunker+Cally2006,
Cameron+etal2008, Moradi+Cally2008, Moradi+etal2009, Felipe+etal2011,
Felipe+etal2010b, Felipe2012, Khomenko+Cally2011, Khomenko+Cally2012}.

Observations of waves in quiet regions have achieved important advances as
well. Multi-line multi-layer observations have put in evidence how the
propagation of waves in plage and network areas, and even in the quiet Sun,
is influenced by the magnetic field (see, e.g., \cite{Krijer+etal2001,
DeMoortel+etal2002, DePontieu+etal2003, Finsterle2004a, Finsterle2004b,
Centeno+etal2009, Vecchio+etal2007, Vecchio2009, Stangalini2011,
Stangalini2012, FujTsun09, MartinezGonzalez2011, Kontogiannis2010a,
Kontogiannis2010b, Kontogiannis2011, Rajaguru2012, Chitta2012}, and
references therein). Unlike in sunspots, magnetic field in the quiet regions
is generally weaker, less organized, and, as a consequence, is more difficult
to measure \cite{SanchezAlmeida2011}. The question of the coupling between
the different solar layers by means of waves, propagating in quiet Sun's
structures, is particularly interesting. Waves with photospheric
periodicities are systematically observed in the solar corona
\cite{DeMoortel+Nakariakov2012}. Nevertheless, it is still an open question
how to get the $p$-mode energy efficiently up there because of a set of
obstacles that waves encounter on their way up, such as: cut-off layer, wave
speed gradients, transition region, etc (see, e.g. \cite{Newington+Cally2010,
Hansen+Cally2012}).

Yet another interesting topic is the practical application of the theoretical
knowledge to infer information about the wave modes from observations.
Diagnostics based on polarimetry or filter imaging have their advantages and
shortcomings, and the information is often masked by radiative transfer
effects and limited spatial and temporal resolution
\cite{Ruedi+Solanki+Bogdan+Cally1999, Norton+Ulrich+Bush+Tarbell1999,
norton01, BellotRubio+etal2000, Settele2002, Khomenko+etal2003}.
Analytical/numerical modeling and Stokes diagnostics should be applied hand
by hand to disentangle information about the wave modes and calculate
wave energy fluxes, or to infer the parameters of magnetic structure, for
example.

All these topics are discussed below with an emphasis on wave propagation in
network, plage and facular regions and their interaction with magnetic
elements present there, leaving aside sunspot waves. Both observational and
theoretical/numerical aspects are highlighted.

\section{Observations of waves in quiet Sun's magnetic structures}
\vspace{0.3cm} \label{sect:obs}

Magnetic elements in the quiet Sun are extremely complex and dynamic. Weak
inter-network magnetic elements move, merge and cancel out, and small scale
loops emerge on granular scale \cite{SanchezAlmeida2004, Centeno2007,
Ishikawa2008, Ishikawa2009, BelloGonzalez2008, OrozcoSuarez2008,
MartinezGonzalez2009, Riethmuller2010, Palacios2012, SanchezAlmeida2011}.
Stronger magnetic elements in the network are generally more organized. Some
recent works were able to resolve a network tube structure, including its
canopy, from spectropolarimetric observations by Sunrise/IMaX \cite{Lagg2010,
MartinezGonzalez2012}. These real tubes are dynamic and asymmetric
structures, unlike the ``canonical'' axisymmetric flux tubes used in many
theoretical wave studies. Locally, acoustic $p$-modes and convective motions
can drive waves in magnetic concentrations present over the quiet solar
surface.

\subsection{Network and inter-network regions}
\vspace{0.2cm} \label{sect:network}

It has long been know that waves of 3-minute periodicity dominate quiet
internetwork chromospheric regions \cite{Deubner+Fleck1990, Lou1995,
Hoekzema+Rutten1998, Rutten+Uitenbroek1991, Lites1993}. These waves have been
observed as bright intermittent \CaIIH$_{2V}$ and $K_{2V}$ grains and are
recognized to be acoustic shocks \cite{Carlsson+Stein1997}.

At the bright network cell borders, long period oscillations of 5--20 min are
usually detected in the photosphere.  These oscillations maintain their
periodicity in the chromosphere and at larger heights \cite{Lites1993,
Krijer+etal2001, DePontieu+etal2003, Bloomfield+etal2006, Tritschler2007,
Vecchio+etal2007}. Shock occurrence is lower at the network cell borders
compared to cell centers and the amplitudes remain below 2--4 km s$^{-1}$ in
the chromosphere.


Three-minute chromospheric oscillations inside supergranular network cells
are well correlated with the underlying photospheric oscillations
\cite{Deubner+Fleck1990, Lites1993}. It is not clear if the same happens to
the 5-minute oscillations at the bright cell borders. Some earlier studies
report that the correlation is difficult to trace \cite{Lites1993}, while
others find a good correlation between the signals at two heights, suggesting
nearly vertical wave propagation from the photosphere to the chromosphere in
the proximity of the network magnetic elements \cite{Lites+etal1982a,
Deubner+Fleck1990, Vecchio+etal2007}. Larger phase shifts between velocity
signals at different heights, $\phi_{VV}$, are found for waves at the network
borders compared to cell centers \cite{Lites+etal1982a, Deubner+Fleck1990}.

\begin{figure*}
\vbox{ \hbox{ \hglue -0.6cm
\includegraphics[width=5.5cm,height=5.5cm]{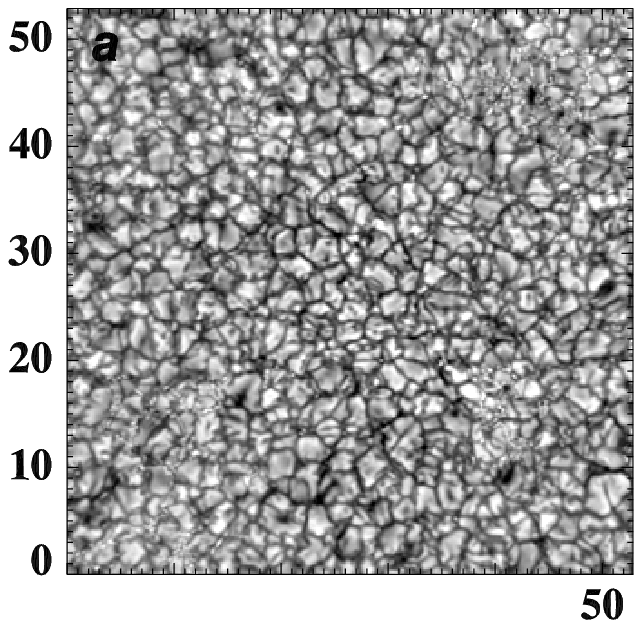}
\hglue -2.15cm
\includegraphics[width=5.5cm,height=5.5cm]{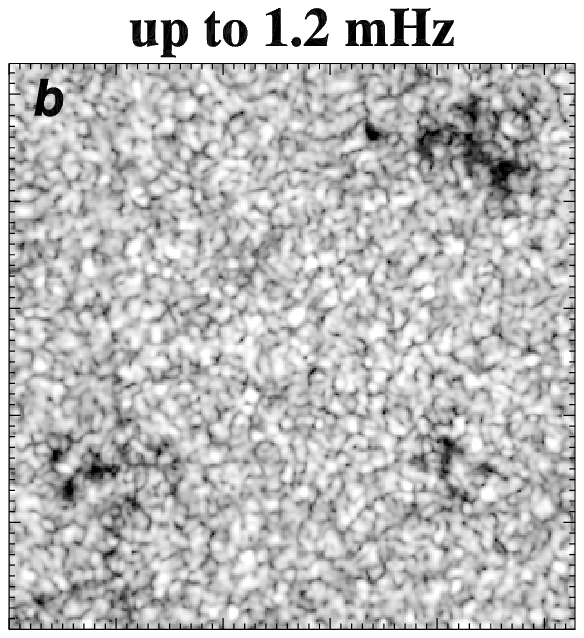}
\hglue -2.15 cm
\includegraphics[width=5.5cm,height=5.5cm]{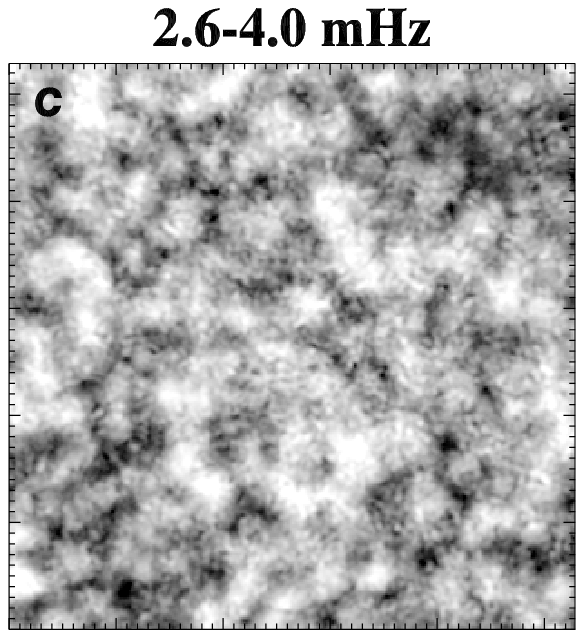}
\hglue -2.15 cm
\includegraphics[width=5.5cm,height=5.5cm]{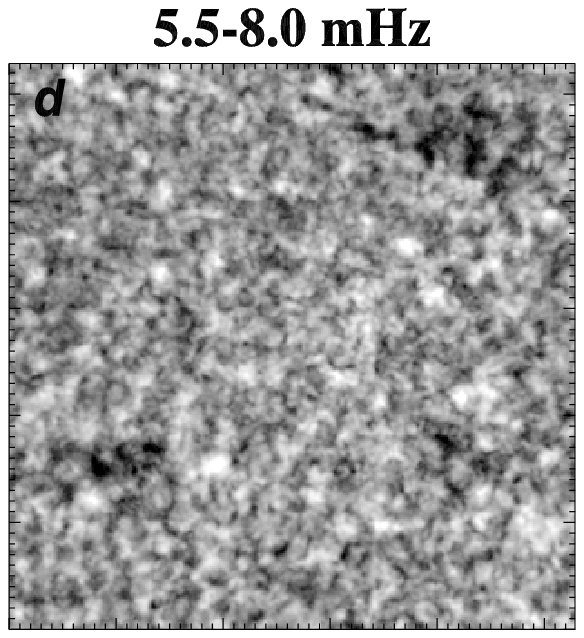}
} \vglue -2.1cm \hbox{
\hglue -0.6cm
\includegraphics[width=5.5cm,height=5.5cm]{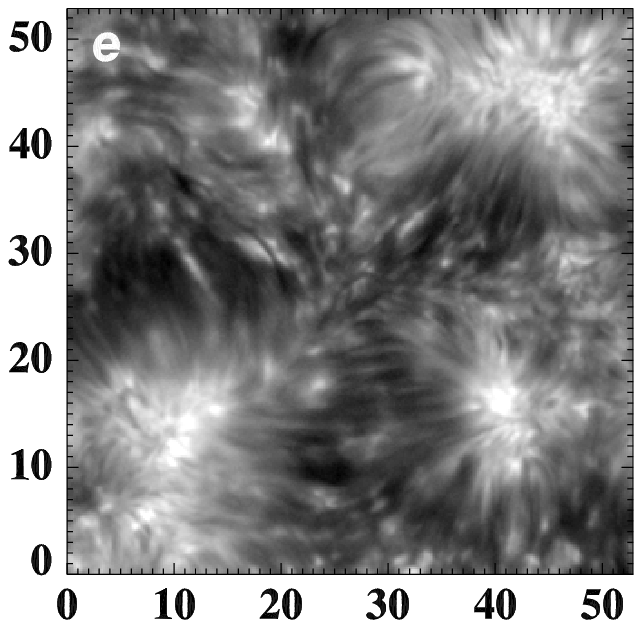}
\hglue -2.15cm
\includegraphics[width=5.5cm,height=5.5cm]{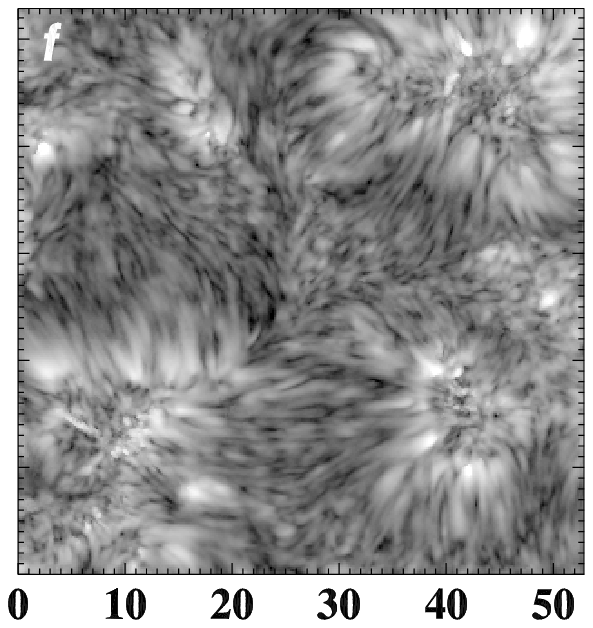}
\hglue -2.15cm
\includegraphics[width=5.5cm,height=5.5cm]{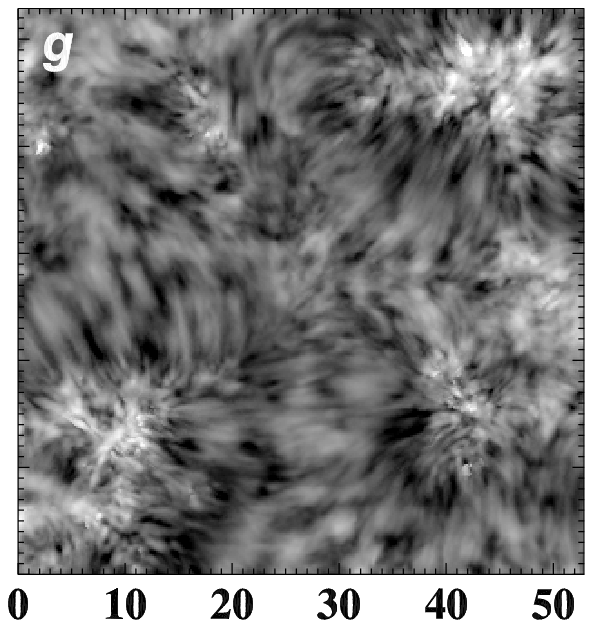}
\hglue -2.15cm
\includegraphics[width=5.5cm,height=5.5cm]{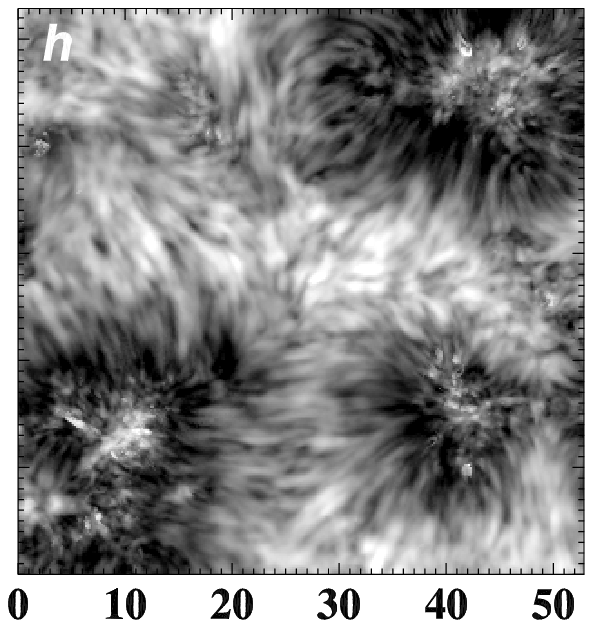}
} } \caption{ Example of IBIS  observations of magnetic shadows from
\cite{Vecchio+etal2007}. {\bf a}: Speckle reconstructed broadband continuum
at 710 nm for the FOV under analysis. {\bf e}: Line core intensity of CaII
854.2 nm. {\bf b--d} and {\bf f--h}: Spatially resolved Fourier's velocity
power maps, averaged over the range of frequencies indicated. The intensity
scale is logarithmic, with bright indicating larger Fourier amplitudes. {\bf
b--d}: photospheric Fe I 709.0 nm line. {\bf f--h}: chromospheric Ca II 854.2
nm line. {\it Courtesy of G. Cauzzi.}} \label{fig:ibis}
\end{figure*}

Recent high-resolution observations from space and ground-based instruments
revealed a much richer behavior of network oscillations related to the
magnetic field topology. It was discovered that short-period waves (3 min
range, $\nu=5-8$ mHz) propagate from the photosphere to the chromosphere only
in restricted areas of the network cell interiors, and that  spatial
distribution of 3-minute chromospheric shocks is highly dependent on the
local magnetic topology \cite{Vecchio+etal2007, Vecchio2009}.
Network magnetic elements are surrounded by ``magnetic shadows'' where the
power of short-period waves is reduced \cite{Krijer+etal2001, Judge+etal2001,
Finsterle2004a, Finsterle2004b, Reardon2009}. An example of this result
borrowed from \cite{Vecchio+etal2007} is given in Figure~\ref{fig:ibis}. A
prominent feature of these observations is the fine structuring of power maps
at different frequencies, following the structure of chromospheric mottles.

Long-period waves (5 min range, $\nu=1.2-4$ mHz) propagate efficiently to the
chromosphere in the close proximity of the magnetic network elements, forming
{\it enhancements} on the power maps (``power halos'')
\cite{Kontogiannis2010a}. These long-period network halos are most prominent
in the photosphere, but are also present in the chromosphere; and are
observed to be co-spatial with chromospheric ``magnetic shadows'' for 3 min
waves.
The halo/shadow areas show positive $\phi_{VV}$ phase shifts indicative of
upward propagating waves at 3 min periodicity, but a mixture of positive and
negative phase shifts for larger-period waves of 5-7 min, with a good
coherence in all cases.

The peculiar power distribution in halos/shadows is apparently linked to the
height of the magnetic canopy, defined as a layer where the plasma $\beta=1$
\cite{McIntosh+Judge2001, McIntosh+etal2001, Kontogiannis2010b,
Kontogiannis2011}.
At the close proximity of network elements, where the canopies are low-lying
(below 1600 km), the {\it chromospheric} power of short-period (3 min) waves
was found to be suppressed, and the {\it photospheric} power of long-period
waves (5-7 min) was found to be enhanced \cite{Kontogiannis2010b,
Kontogiannis2011}. It was suggested that the 5-7 min power enhancements in
the photosphere may be due to reflection of waves on the overlying canopy.
These enhancements are observed at all magnetic field inclinations, but are
specially significant for more inclined canopy fields at the close proximity
of magnetic elements, suggesting more efficient mode conversion there
\cite{Kontogiannis2010b}.

Table~\ref{tab:network} summarizes the properties of oscillations in the
quiet network areas, in terms of their periods.

 \begin{table}[b]
 \begin{center}
 \caption[]{\label{tab:network}
           {Periods of network \& inter-network oscillations }}
           \vspace{0.1cm}
\begin{tabular}{llll}
\hline
               & Centers of      &  Close          & Inter-network \\
               & mag. elements   &  surroundings   & beyond mag. elements        \\
\hline
\vspace{0.1cm}
  Photosphere  & 5 min           & 5 min           & 5 min        \\
  Chromosphere & Possibly 5 min  & 5 min enhanced  & 3 min shocks \\
\vspace{0.1cm}
               &                 & 3 min shadowed  &              \\
\vspace{0.1cm}
  Propagation  & Up \& down      & Up \& down      & Up           \\
  \hline
\end{tabular}
\end{center}
\end{table}

 \begin{table}[b]
 \begin{center}
 \caption[]{\label{tab:plage}
           {Periods of facula \& plage oscillations }}
           \vspace{0.1cm}
\begin{tabular}{llll}
  \hline
               & Centers of             &  Close        & Halo  \\
               & mag. elements          &  surroundings & areas       \\
\hline
  Photosphere  & 5 min damped           & 5 min damped      & 3 min enhanced \\
\vspace{0.2cm}
               & 3 min enhanced         &  3 min enhanced   &                \\
\vspace{0.2cm}
  Chromosphere & Not clear              & 5 min for inclined $B$ & 3 min enhanced   \\
\vspace{0.2cm}
  Propagation  & Possibly up            & Possibly up    & Up and down \\
\hline
\end{tabular}
\end{center}
\end{table}

\subsection{Plage and facular regions}
\vspace{0.2cm}  \label{sect:plage}

Plage and facular regions contain larger magnetic flux structures compared to
the network regions discussed above. Early works (based on slit spectra)
showed that the amplitude of 5 min oscillations in the photosphere decreases
by about 25\% in plage regions with an average flux above 80--100 G, compared
to the quiet Sun, while shorter-period oscillations strengthen
\cite{Deubner1967, Howard1967, Blondel1971, Woods1981}. More recent
spectropolarimetric observations confirm that the amplitude of 5-min
oscillations decreases by 20-40\% in the facula photosphere
\cite{Kobanov+Pulyaev2007}. These observations also reveal that the power of
5-min oscillations increases significantly in the chromosphere, dominating
the power spectrum \cite{Kobanov+Pulyaev2007, Centeno+etal2009}. The phase
spectra of the 5-min facular waves are compatible with the vertical wave
propagation from the photosphere to the chromosphere.
One may conclude then, that the dominance of 5-min periodicities in the
chromosphere is a common feature of both network and plage waves. However,
the picture does not seem to be that simple.

The 2D filtergram observations show more pieces of the puzzle regarding the
power distribution for long-period waves in relation to the magnetic field
topology. It was argued that inclined magnetic field lines at the boundaries
of large-scale enhanced network cells provide ``portals'' through which
long-period (5 min) waves can propagate into the solar chromosphere, due to
the lowering of the cut-off frequency in the low-$\beta$ environment
\cite{DePontieu+etal2004, DePontieu2005, Hansteen2006, Jefferies+etal2006,
McIntosh+Jefferies2006}. Larger propagation speeds of chromospheric waves
were indeed found for strong fields inclined more than 40 degrees. At the
center of plage magnetic elements (where the field is expected to be close to
the vertical) preferably 3 min oscillations are observed to be transmitted to
the chromosphere, while 5 min oscillations were transmitted at the edges of
plage magnetic elements, where the field expands and becomes inclined
\cite{Wijn2009}.

The 2D filtergram observations rely on potential extrapolations from
photospheric longitudinal magnetograms, unlike spectropolarimetric
observations where the inversion of Stokes profiles gives information about
the photospheric magnetic field vector (see e.g. \cite{Centeno+etal2009}),
suggesting it to be essentially vertical and thus giving strong arguments for
the vertical propagation. On the other hand, 2D filtergrams have much higher
resolution and field of view needed to detect the inclined propagation, if it
takes place.
The advantages of both observing techniques (imaging and spectropolarimetry)
were exploited in \cite{Stangalini2011, Stangalini2012}, by using
simultaneous IBIS photospheric and chromospheric measurements together with
spectropolarimetric inversion of the SOT/Hinode data. For a plage region
containing a pore, maximum transmission of 3 min oscillations was found for
field inclined by 15 degrees, while 5 min oscillations were transmitted for
the field inclined by 25 degrees, providing support for the inclined
propagation of long-period waves, at least for the strong field of a pore. It
is not clear, however, if the same happens for the weaker magnetic elements,
and more data as those by \cite{Stangalini2011, Stangalini2012} are clearly
needed.

For shot-period waves (3 min range, $\nu=5.5-7.5$ mHz), the power
distribution over plage regions shows a suspicious enhancement, both in the
photosphere \cite{Brown+etal1992} and in the chromosphere
\cite{Braun+etal1992, Toner+LaBonte1993}. These power enhancements are known
as ``halos'' and have been much discussed in the literature on local
helioseismic waves\footnote{Note that halos in plage are observed at
short-periods, unlike long-period power halos surrounding network magnetic
elements, discussed in Section~\ref{sect:network}.}. The acoustic power
measured in halos is higher than in the quite Sun by about 40-60\%
\cite{Hindman+Brown1998, Braun+Lindsey1999, Donea+etal2000, Jain+Haber2002,
Nagashima+etal2007}. The halos are observed at longitudinal magnetic fluxes
$\langle B \rangle= 50-300$ G \cite{Hindman+Brown1998,
Thomas+Stanchfield2000, Jain+Haber2002}. In the photosphere the halos are
located at the edges of plage regions, while in the chromosphere they extend
to a large portion of the nearby quiet Sun \cite{Brown+etal1992,
Braun+etal1992, Thomas+Stanchfield2000}. The characteristic period of maximum
enhancement decreases with field strength and the wave spectrum in halos is
shifted to higher wave numbers at constant frequency \cite{Gizon2009,
Schunker2011}. The wave power in halos is observed to be modulated with the
field inclination, being stronger for more inclined fields
\cite{Schunker2011}. The height-dependent changes in halo properties seen in
HMI/AIA SDO data provide signatures of magneto-acoustic wave refraction
sensitive to magnetic field strength and topology \cite{Rajaguru2012} (the
presence of up and down-going waves has also been reported previously by
\cite{Braun+Lindsey2000b}).



The vast amount of information on the observed halo properties is mostly
obtained from the observational datasets typical for local helioseismology
(i.e. long time series over a large portion of the Sun, including a complete
active region, but medium spatial resolution, as SOHO/MDI or SDO/HMI). It is
not clear how to consistently compare these large-scale data with higher
resolution (but smaller area) IBIS, SST and Hinode data
\cite{DePontieu+etal2003, Hansteen2006, Wijn2009, Stangalini2011,
Stangalini2012} or chromospheric/transition region TRACE data
\cite{DePontieu+etal2004, DePontieu2005, Jefferies+etal2006,
McIntosh+Jefferies2006}, resolving propagation in individual magnetic
structures. In addition, there is much confusion about the waves observed in
plage, facular, or enhanced network regions, often discussed in the same
context. It is not evident beforehand if the observed wave behavior should be
the same, since the field strength of the magnetic structure definitely plays
a crucial role.

Table~\ref{tab:plage} makes an attempt to summarize the properties of
oscillations in plage areas, though definite conclusions are hard to derive
in some cases.

\begin{figure*}[!t]
\centering{
\includegraphics[width=0.16\textwidth, bb= 116 400 418 702]{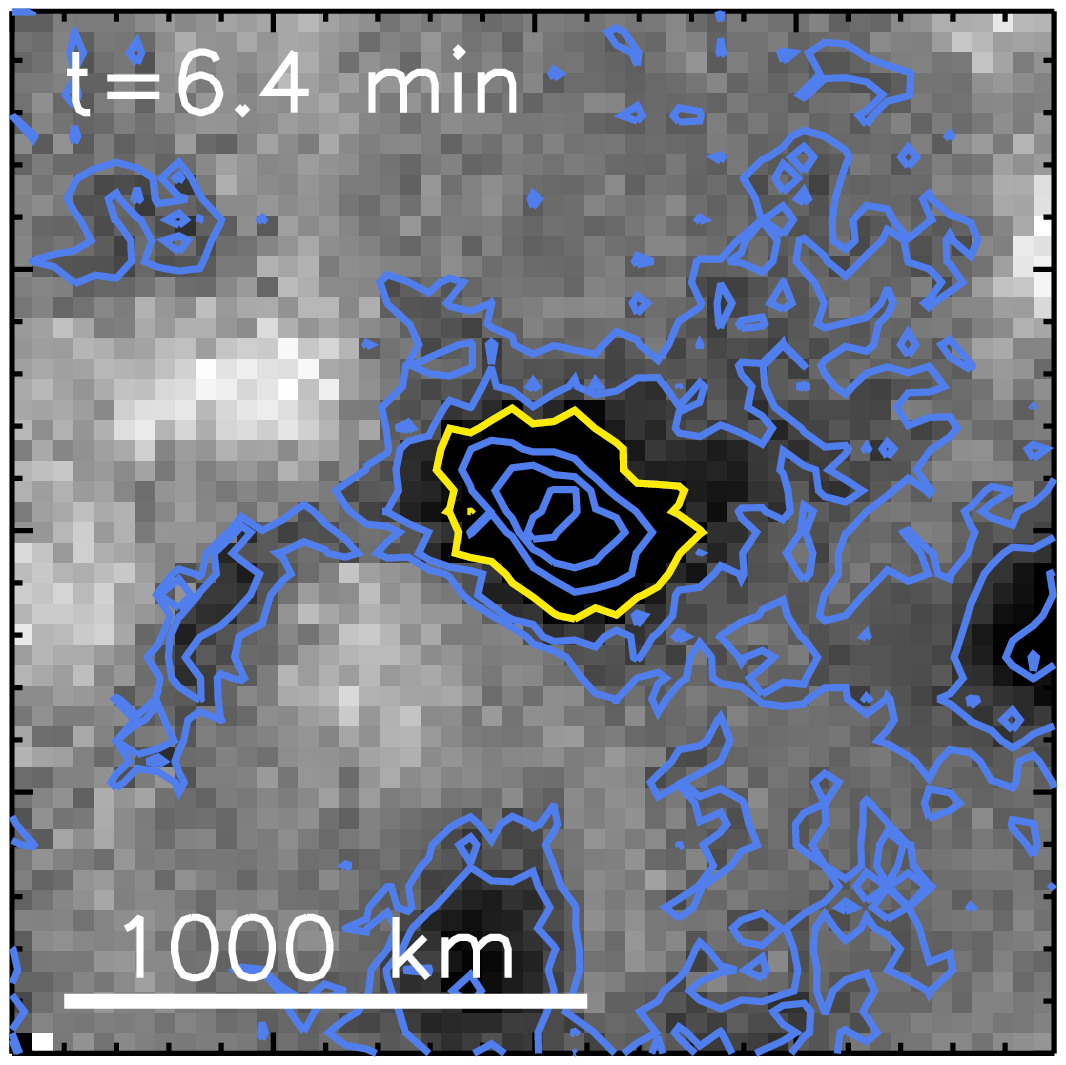}
\hspace{-0.15cm}
\includegraphics[width=0.16\textwidth, bb= 116 400 418 702]{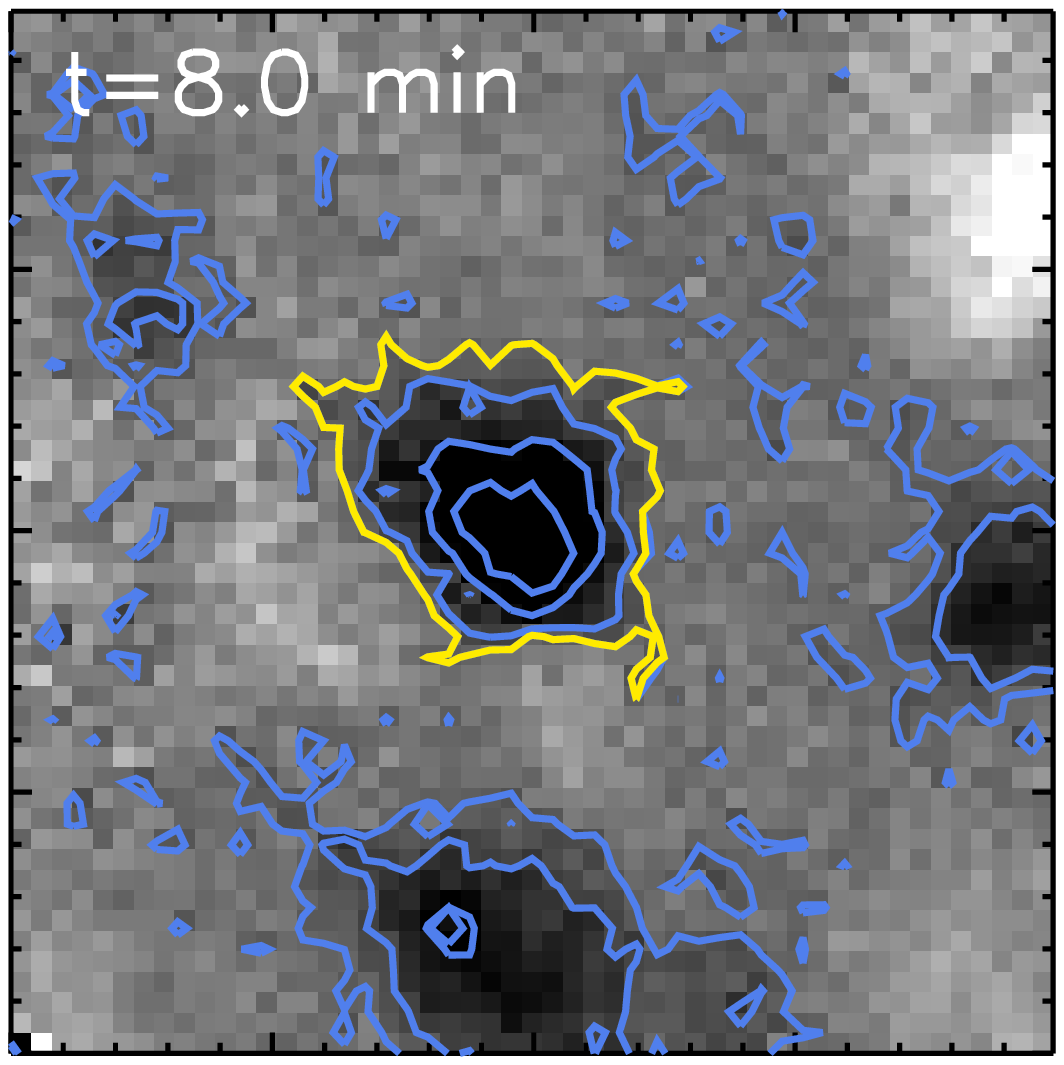}
\hspace{-0.15cm}
\includegraphics[width=0.16\textwidth, bb= 116 400 418 702]{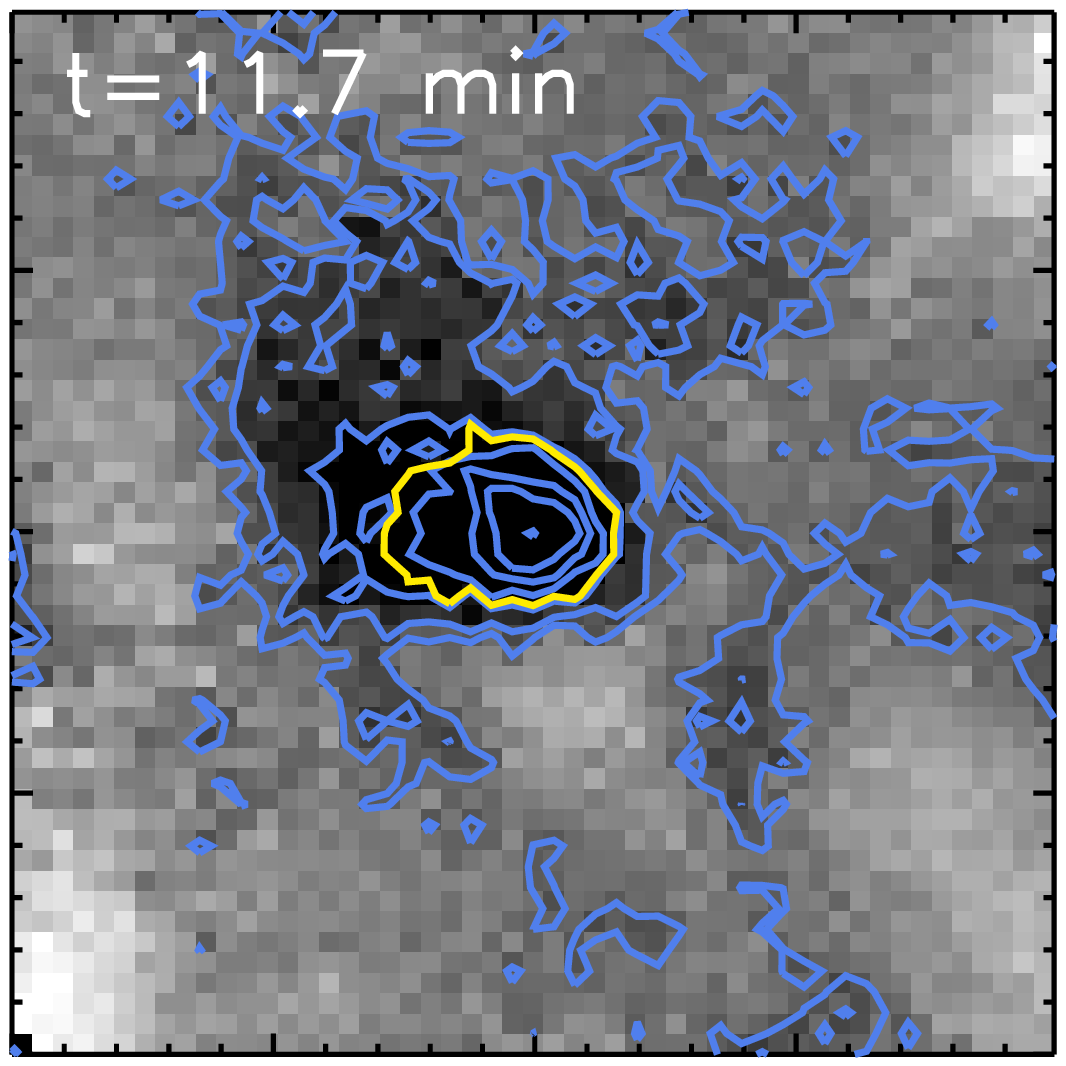}
\hspace{-0.15cm}
\includegraphics[width=0.16\textwidth, bb= 116 400 418 702]{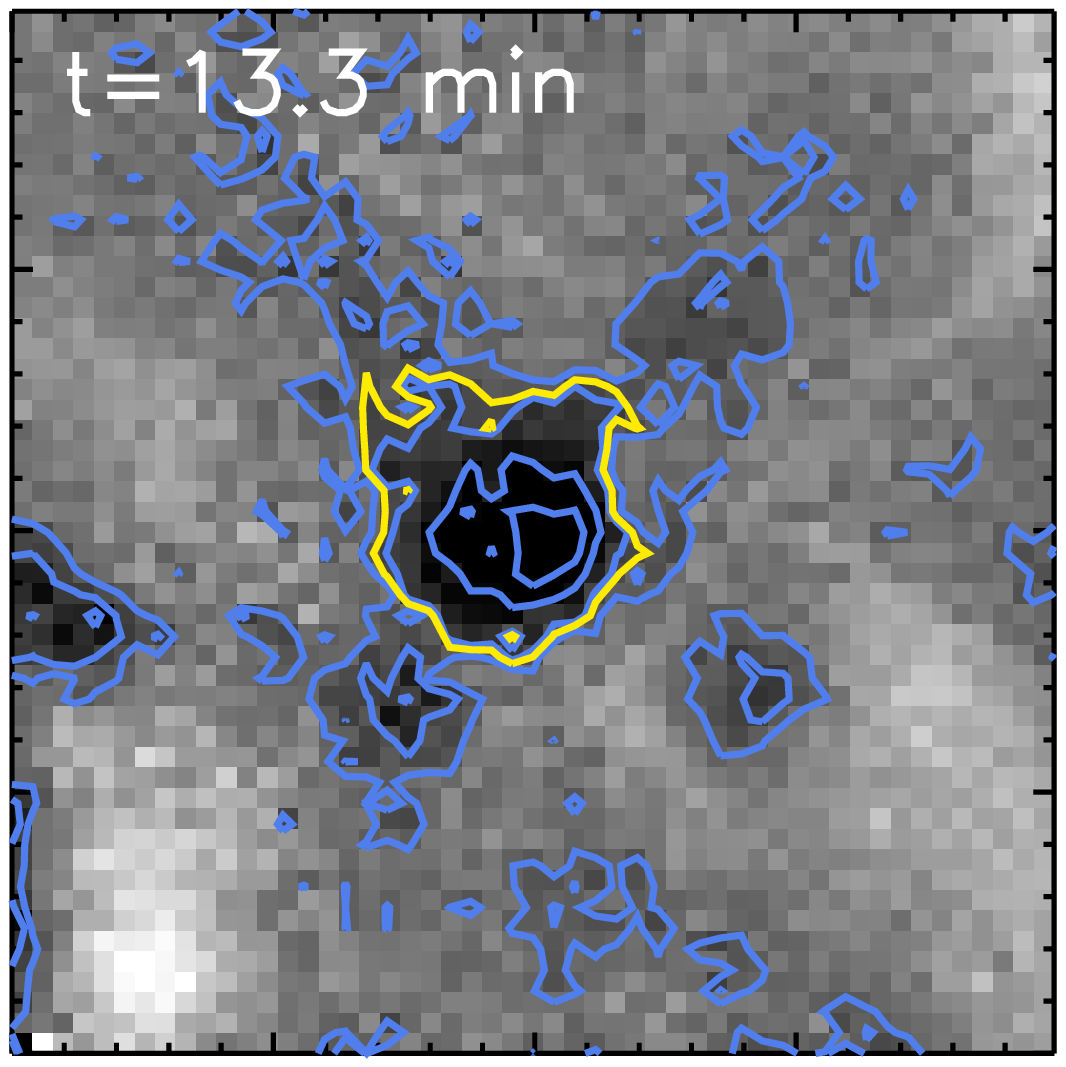}
\hspace{-0.15cm}
\includegraphics[width=0.16\textwidth, bb= 116 400 418 702]{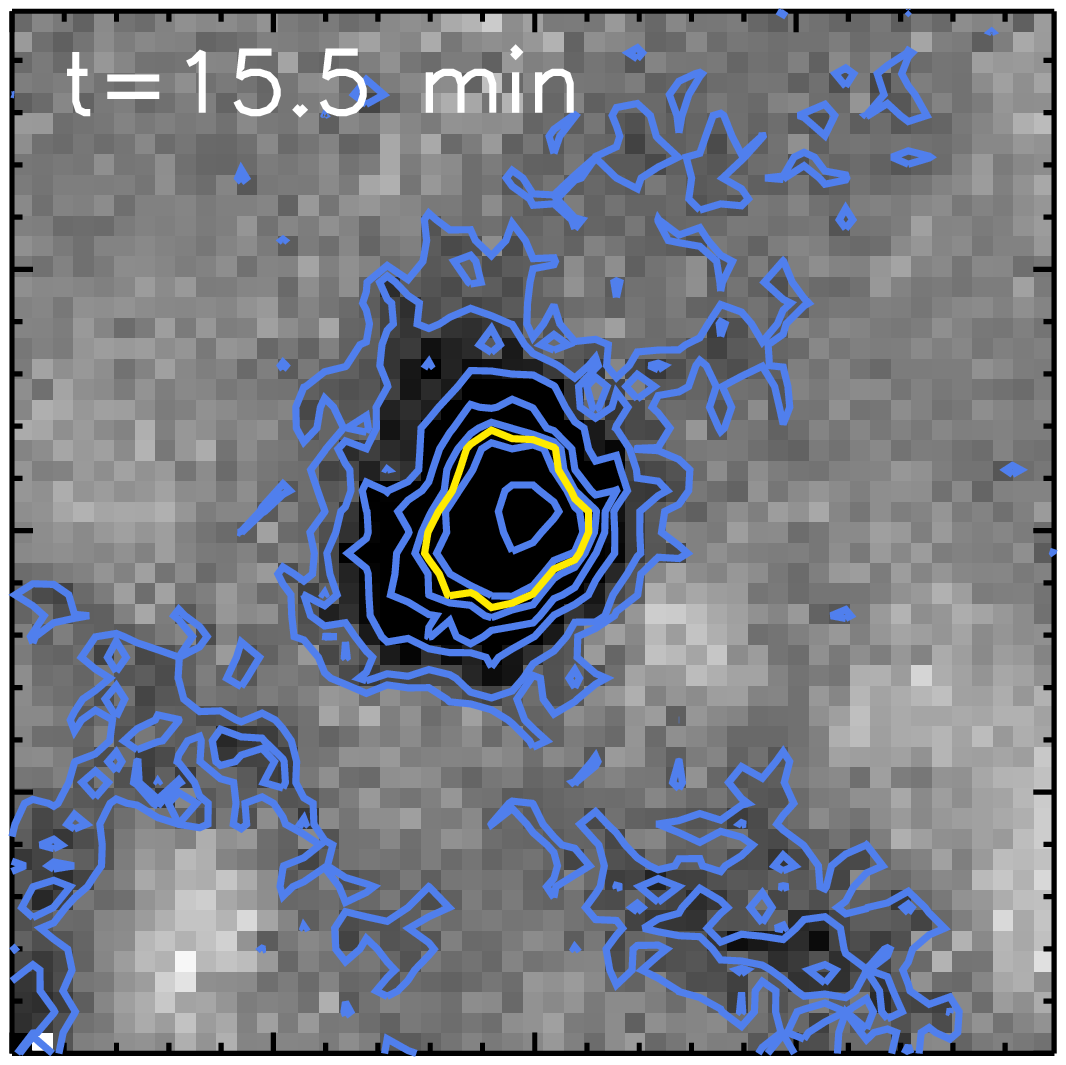}
\hspace{-0.15cm}
\includegraphics[width=0.16\textwidth, bb= 116 400 418 702]{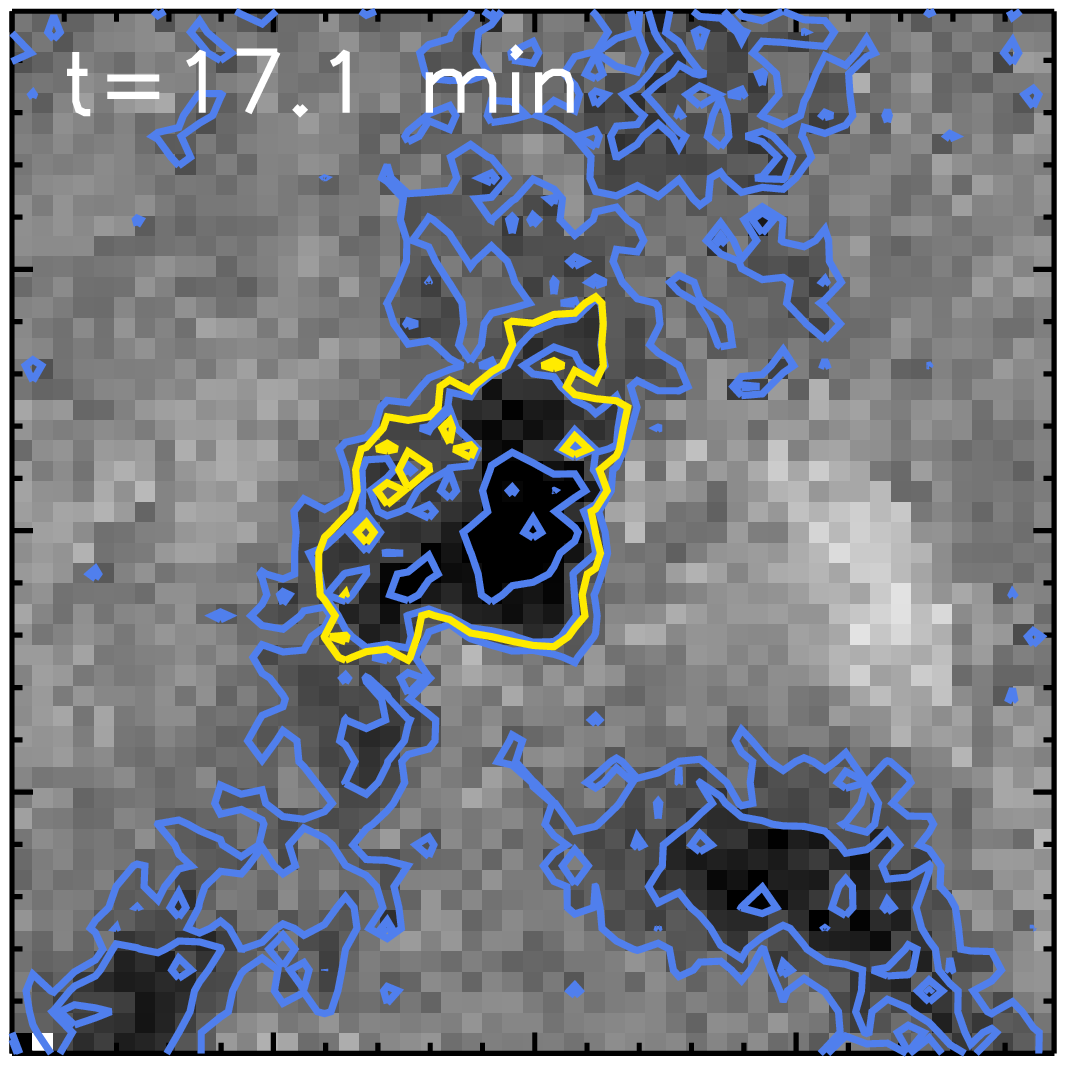}}
\caption{Example of time evolution of a weak circular polarization patch in
the quiet Sun from \cite{MartinezGonzalez2011}. The black and white
background represents the magnetic flux density computed in the weak-field
approximation, saturated to $\pm$20 Mx cm$^{-2}$.
Yellow line is the iso-magnetic flux
density contour containing a time-constant magnetic flux of -5$\times
10^{16}$ Mx. {\it Courtesy of M. Mart{\'{\i}}nez Gonz{\'a}lez.}}
\label{fig:salchicha}
\end{figure*}

\subsection{Magnetic field oscillations}
\vspace{0.2cm}

Up to here, we have discussed oscillations of velocity and intensity. Magnetic
field measurements, when performed, were mostly done to obtain the magnetic
topology, but not oscillations of the magnetic flux (or magnetic field
vector) itself.

Even in strong sunspot regions it has been hard to detect magnetic field
oscillations due to the smallness of the expected amplitudes and due to
radiative transfer and instrumental effects, masking the real oscillations
\cite{Ruedi+Solanki+Stenflo+Tarbell+Scherrer1998,
Ruedi+Solanki+Bogdan+Cally1999, norton01, Settele2002, Khomenko+etal2003,
Kobanov+Pulyaev2007}.

The detection of magnetic flux/field oscillations due to magneto-hydrodynamic
waves in the quiet regions has been elusive. The relations between the
magnetic flux, velocity and intensity oscillations in the photosphere of
pores and intergranular magnetic elements, obtained from homogeneous,
seeing-free SOT/Hinode data, reveal root-mean-square amplitudes of magnetic
flux fluctuations of 4$-$17 G (0.3\%$-$1.2\% from the background value),
corresponding to velocity oscillations of $\sim$100 m s$^{-1}$ and intensity
oscillations of 0.1\%$-$1\% \cite{FujTsun09}.  Velocity was leading the
magnetic flux oscillations by a quarter of cycle, and magnetic flux
oscillations were 180 degree out of phase with the intensity oscillations.
These phase lags were interpreted as due to superposition of the ascending
and the descending sausage and kink mode waves reflected at the transition
region \cite{FujTsun09}.

Another recent detection of  oscillations of magnetic flux density in quiet
Sun magnetic elements was done from high-resolution Sunrise/IMaX data
\cite{MartinezGonzalez2011}. It was discovered that the area of patches with
constant magnetic flux oscillates in response to granular forcing in a range
of periods similar to granular life times (implying that magnetic field
strength oscillates in antiphase). An example of such observation borrowed
from this work is shown in Figure~\ref{fig:salchicha}. Curiously, the
amplitude of these area oscillations is highly non-linear with more than
100\% variations from the mean value. No apparent correlation with velocity
or intensity oscillations was found.

\section{Models}
\vspace{0.3cm}

This section gives an overview on recent, mostly numerical, models of the
wave propagation in solar magnetic structures. We subdivide the simulations
into several (overlapping) groups, one of them aiming to explain the power
distribution in and around magnetic features, another group studying the
energy propagation by waves from the photosphere to the chromosphere,
transition region and corona, yet another one for the mode conversion.

\subsection{Coupling of the photosphere, chromosphere and above}
\vspace{0.2cm}

Small-scale magnetic field concentrations are ubiquitous in the quiet Sun and
magnetic waves driven in these elements by $p$-mode  or granular forcing are
obvious candidates to transport energy to heat the upper layers. Solar
magnetic elements are far more complex than canonical axisymmetric flux
tubes. But despite the clear simplifications of the tube structure,
theoretical works on propagation of waves in flux tubes are of great
importance to understand the basic physics of the wave mode behavior in small
scale magnetic elements.
There have been numerous analytical works using the thin flux tube
approximation, when the radius of the tube is considered to be much smaller
than the local vertical scale height, and the internal flux tube structure
can be neglected \cite{Roberts+Webb1978, Spruit1981, Roberts1981,
Roberts1983, Cally1986, Ruderman2002, Dymova+Ruderman2005, Roberts2006,
Diaz+Robertts2006, Ruderman2009} (see \cite{Solanki1993,Solanki2006} for a
review).
In fact, the comparison to simulations of magneto-convection suggests that a
second-order thin flux tube approximation might do reasonably well for
super-equipartition magnetic field concentrations like those existing in the
simulations \cite{YellesChaouche2009}. This argument can be used for wave
studies if asymmetry and flows are not crucial for the considered problem.

The number of idealized simulations of flux tube waves propagating from the
photosphere to the chromosphere and above is ever increasing during the last
several years \cite{Hasan2000, Hasan+etal2003, Hasan+Ulmschneider2004,
Hasan+etal2005, Hasan+Ballegooijen2008, Erdelyi2007, Khomenko+etal2008a,
Vigeesh2009, Vigeesh2011, Vigeesh2012, Nutto2010, Fedun2011a, Fedun2011b,
Kato+etal2011}. Different ways of driving flux tube waves have been checked:
horizontal motions producing transverse kink waves; pressure fluctuations
producing longitudinal waves; twisting motions generating torsional Alfv\'en
waves; observationally-driven or impulsive motions. The general idea behind
all these works is to understand how much of the wave energy can reach the
chromosphere and corona, and whether the wave modes reaching there can be
easily dissipated to convert their energy into heat (as, e.g. compressible
acoustic modes).

\begin{figure*}
\begin{center}
\includegraphics[width=14cm]{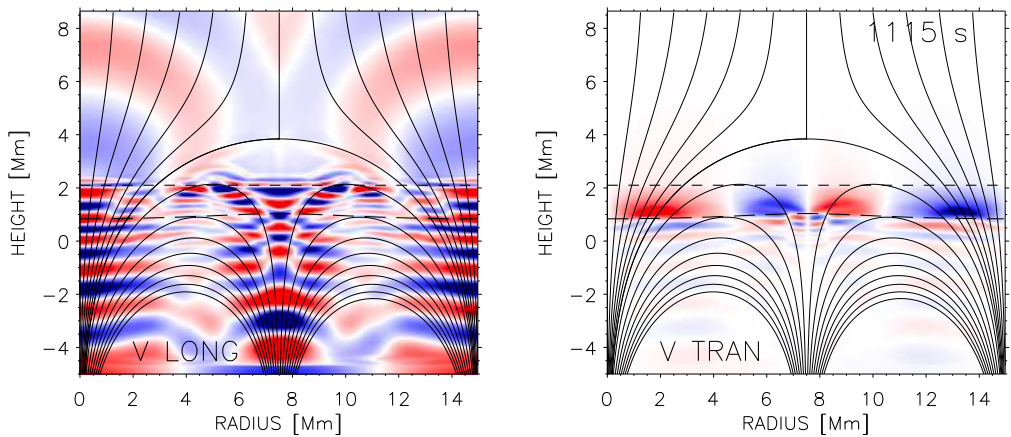}
\end{center}
\caption{Snapshot of longitudinal (left) and transversal (right) velocities,
scaled by a factor or $\sqrt{\rho c_s}$ and $\sqrt{\rho v_a}$,
correspondingly, in the 2D simulations of fast (acoustic) mode propagation
and conversion from sub-photosphere to the low corona from
\cite{CalvoSantamaria2013}. } \label{fig:arco}
\end{figure*}

The transverse foot-point driving was extensively studied in a series of
papers \cite{Hasan+etal2003, Hasan+Ulmschneider2004, Hasan+etal2005,
Vigeesh2009, Vigeesh2012} by means of 1D and 2D simulations. These works
mostly aim at explaining waves in bright network elements at cell borders
observed in \CaIIH\ (see Section~\ref{sect:network}) and their relation to
G-band bright points. It was discovered an efficient mechanism of non-linear
coupling between the kink waves (generated by transverse foot point driving)
and longitudinal waves, forming shocks and supplying energy to the
chromosphere. The efficiency of the coupling was found to be a function of
the plasma $\beta$, being maximum for tubes with $\beta=0.2$ when the kink
and tube wave speeds are equal \cite{Hasan+Ulmschneider2004}. Adding a second
dimension to the modeling (2D), the transverse foot point motions were seen
to excite fast and slow magneto-acoustic waves inside the tube, and acoustic
ones at the interface between the tube and the non-magnetic medium
\cite{Hasan+etal2005}. The acoustic waves at the interface steepen to shocks
in the chromosphere and flux tubes with stronger magnetic field produce these
shock waves more efficiently \cite{Vigeesh2009}.

The above works were mostly focused on short-period (10-20 sec) waves or
impulsive driving, and it is hard to make a direct comparison to the
observations. Nevertheless, similar processes of magneto-acoustic mode
generation and transformation were shown to take place when considering more
realistic driving with 3 and 5 min periodicities \cite{Khomenko+etal2008a}.
Complex shock wave patterns are formed across the tube, with shock waves
propagating alongside the tube borders. Sub-arcsec spatial resolution would
be needed to observationally resolve this pattern \cite{Khomenko+etal2009b,
Vigeesh2011}.

Apart from linear and non-linear mode transformation, slow chromospheric
shocks in magnetic flux tubes were demonstrated to be produced by a mechanism
of ``turbulent pumping'' \cite{Kato+etal2011}, i.e. ``massaging'' of the
magnetic tubes in intergranular lanes by external downflows. Possibly, a
similar process is responsible for the periodic area variations of quiet Sun
magnetic elements discovered in \cite{MartinezGonzalez2011}.

When the tube foot points are twisted instead of being transversally shaken,
torsional Alfv\'en waves can be generated. There are indications that the
energy input into the chromosphere is lower for the twisting driver, compared
to the transverse shaking \cite{Vigeesh2012}. The torsional Alfv\'en waves,
generated by twisting driver seem to be frequency-filtered by the tube's
three-dimensional structure, with higher frequencies transmitted at the
central part of the tube, and lower ones at the sides  \cite{Fedun2011a}.


The next challenge for the wave numerical modeling in 2D and 3D is to extend
the analysis into the transition region and corona, in order to explain how
waves with photospheric periodicity leak their energy to the upper
atmosphere. This is numerically challenging, since many pressure scale
heights have to be modeled at the same time, and the wave speeds change
considerably. The transition region represents a discontinuity in the wave
speed and waves are expected to be efficiently reflected there. Thus, it is
unclear how much energy can go through. Nevertheless, short-period (30 sec)
magneto-acoustic waves  were shown to be transmitted rather efficiently
through the transition region, exciting horizontally propagating disturbances
there \cite{Fedun2011b}. Figure~\ref{fig:arco} shows an example of the
modeling of wave propagation from the sub-photosphere to the chromosphere,
transition region and corona for realistic periods in the 3-5 min range
\cite{CalvoSantamaria2013}. It demonstrates a complex wave pattern formed by
multiple mode transformations at the $\beta=1$ layer, wave refractions and
reflections at the transition region and a 2D magnetic null point located
above. More work will be needed in the future to quantify these complex
interactions, and to evaluate the transmitted energy, depending on the
magnetic field configuration and strength.

\subsection{Mode transformation in 2D and 3D}
\vspace{0.2cm}

Linear mode transformation is an important process that enables different
wave modes to exchange their energy when their propagation speeds are close.
Mode transformation between fast and slow magneto-acoustic modes in 2D was
extensively studied analytically and numerically
\cite{Zhugzhda+Dzhalilov1982, Cally+Bogdan+Zweibel1994, Cally+Bogdan1997,
Cally2005, Cally2006, Rosenthal+etal2002, Bogdan+etal2003, Crouch+Cally2003,
Schunker+Cally2006, Khomenko+Collados2006}. A review on these works before
2008 can be found in \cite{Khomenko2009}. In brief, the direction and the
effectiveness of the mode transformation depends on the wave frequency and
the attacking angle between the wave vector $\vec{k}$ and the magnetic field
$\vec{B}$ at the layer of plasma $\beta=1$. The fast-to-slow mode
transformation is complete for waves with $\vec{k}
\parallel \vec{B}$. The fast-mode high-$\beta$ waves
launched from their sub-photospheric lower turning points typically reach the
transformation layer with an angle close to 20--30 degrees, so the
transformation is particularly strong for magnetic fields inclined by this
amount \cite{Crouch+Cally2003, Cally2006, Schunker+Cally2006}.

Recently, mode transformation in 3D has attracted a lot of attention. It
enables to produce Alfv\'en waves from fast magneto-acoustic waves above
their reflection height in the low chromosphere, a mechanism that may allow
to transport wave energy through the transition region to the corona.
Alfv\'en waves are claimed to be detected by numerous observations, but the
discussion on the nature of the detected waves is still ongoing
\cite{De-McICar07aa, TomMcIKei07aa, VanNakVer08aa, Pascoe2010,
DeMoortel+Nakariakov2012}.

\begin{figure*}
\includegraphics[width=5.5cm]{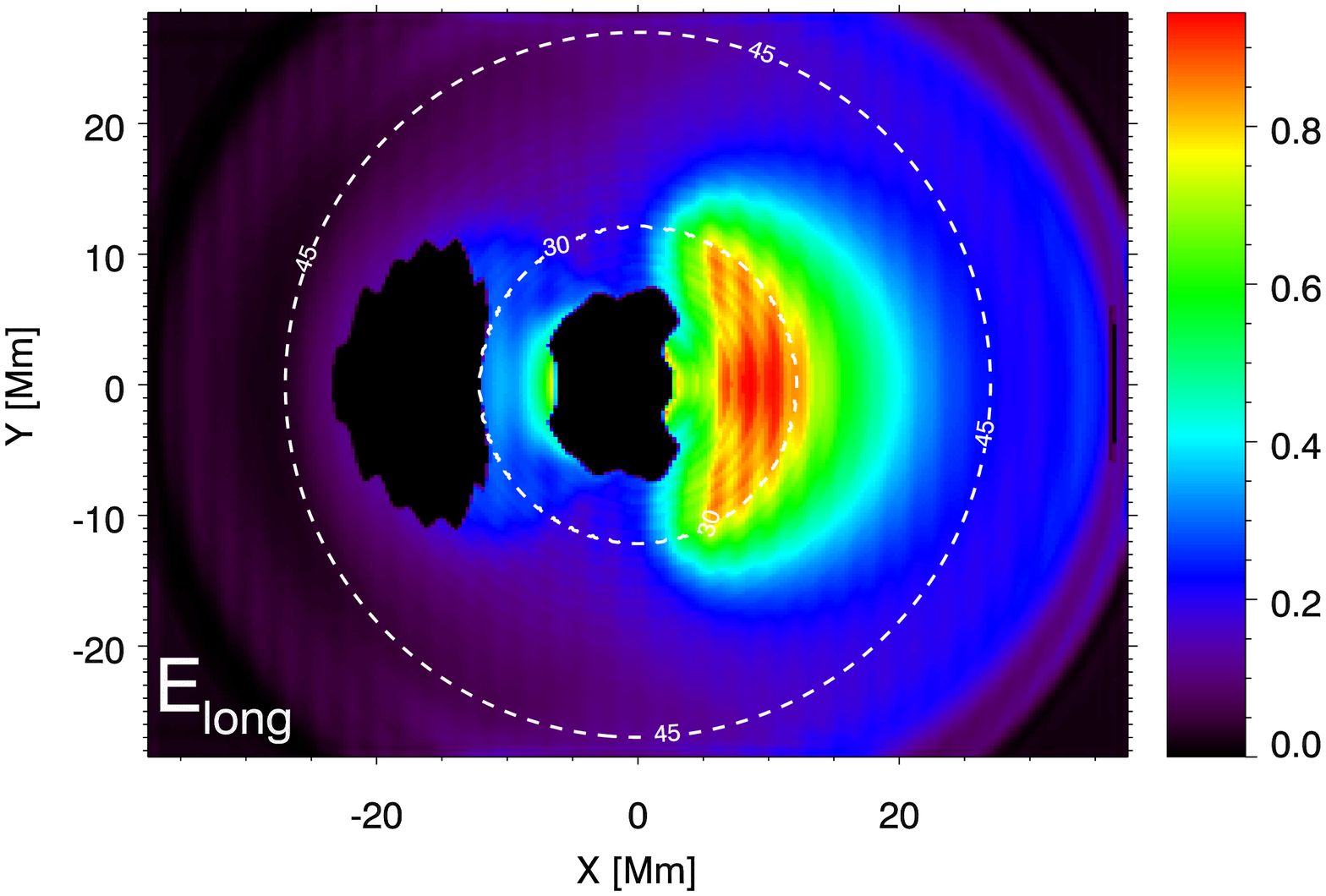}
\includegraphics[width=5.5cm]{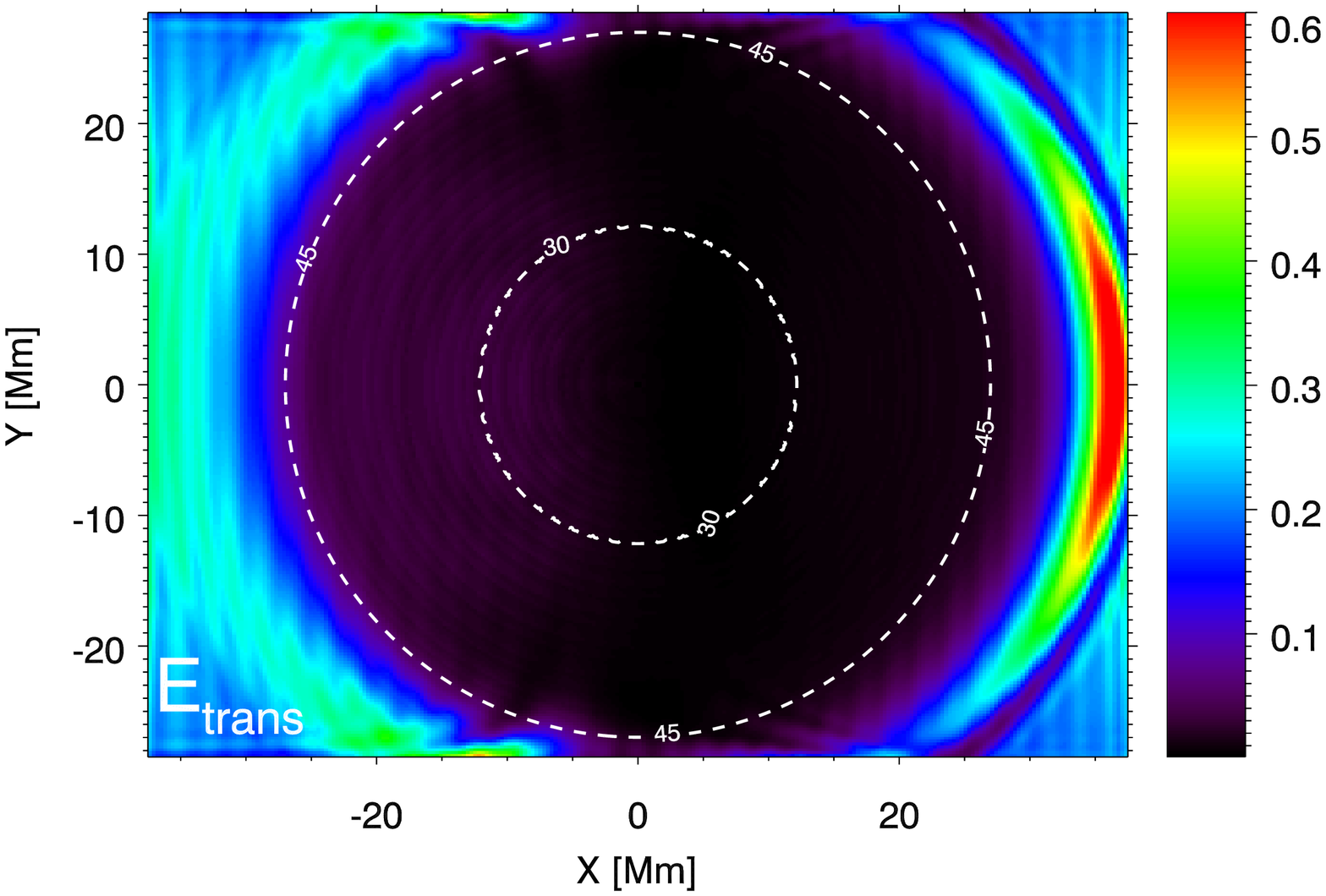}
\includegraphics[width=5.5cm]{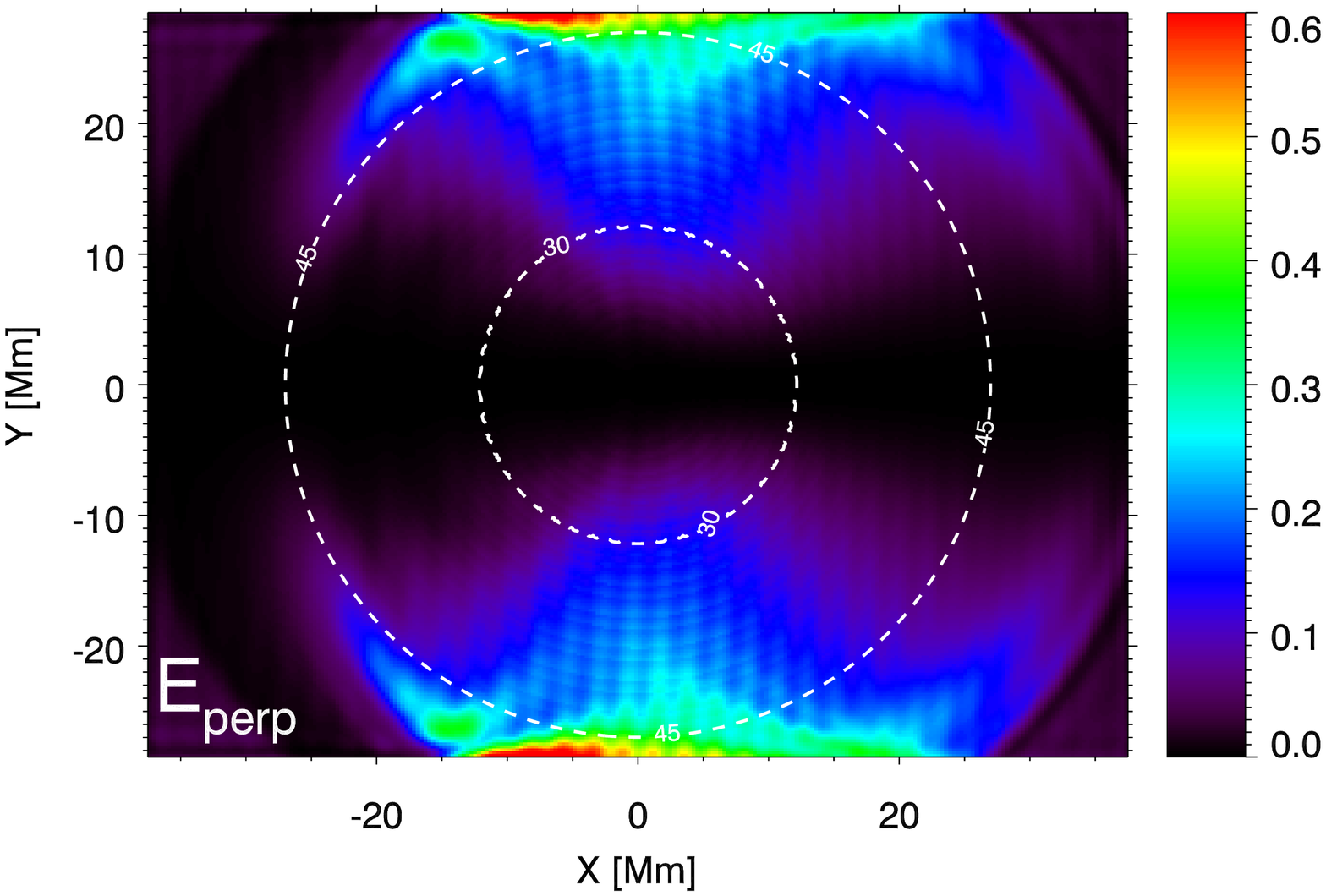}
\caption{Example of wave energy fluxes in the chromosphere of a sunspot model
from \cite{Felipe2012} showing the redistribution of the energy of different
modes due to three-dimensional mode conversion. Left panel: acoustic flux due
to slow longitudinal waves; middle panel: magnetic flux due to fast waves;
right panel: magnetic flux due to Alfv\'en waves. The units of the color
coding are $10^6$ erg cm$^{-2}$ s$^{-1}$. {\it Courtesy of T. Felipe.} }
\label{fig:alfven}
\end{figure*}

The pioneering study of 3D mode conversion from fast to Alfv\'en waves for
homogeneous inclined fields discovered it to be most efficient for preferred
magnetic field inclinations between 30 and 40 degrees, and azimuth angles
between 60 and 80 degrees \cite{Cally+Goossens2008}. This orientation allowed
the best alignment with $\vec{B}$ of the fast mode wave propagating from its
sub-photospheric lower turning point. Alfv\'enic fluxes transmitted to the
upper atmosphere are found to be similar, or larger, than acoustic fluxes at
some orientations. Low-frequency gravity waves can also be converted
efficiently into Alfv\'en waves at large magnetic field inclinations
\cite{Newington+Cally2010}.

While the fast-to-slow mode conversion occurs at the layer where plasma
$\beta=1$, the fast-to-Alfv\'en conversion takes place around and above the
fast wave reflection height $z_{\rm refl}$ in 3D (see Fig. 1 in
\cite{Khomenko+Cally2012}), localized more closely to $z_{\rm refl}$ as the
frequency increases \citep{CallyHansen2011}. For 3-5 min waves, the
conversion process is typically spread over much of the chromosphere. Both
up-going or down-going Alfv\'en waves can be produced depending on the
relative orientation of the wave vector and magnetic field at heights where
the fast mode undergoes the refraction, going up and down
\cite{CallyHansen2011, Khomenko+Cally2012}.

Numerical simulations of conversion to Alfv\'en waves require 3D or, at
least, 2.5D geometries \cite{Felipe+etal2010a, Felipe2012,
Khomenko+Cally2011, Khomenko+Cally2012}. Recent 2.5D simulations show that
the global picture of conversion to Alfv\'en waves remains valid when
considering complex field configurations, appropriate for large-scale sunspot
magnetic structure \cite{Khomenko+Cally2012}. The conversion to Alfv\'en
waves is particularly important for strongly inclined fields like those
existing in sunspot penumbrae. A fully 3D simulation, covering a large
portion of a sunspot model with a wide range of inclinations, nicely
demonstrated that a significant energy flux due to Alfv\'en waves can indeed
be transmitted to the chromosphere at large field inclinations
\cite{Felipe2012}.  The upward-propagating flux due to Alfv\'en waves at the
peripheric parts of the sunspot model was found to be as large as the
acoustic flux due to slow longitudinal waves propagating close to the axis.
An example of this result is shown in Figure~\ref{fig:alfven}.

Inclusion of radiative losses into the model does not change significantly
the wave paths neither the conversion picture \cite{Newington+Cally2011}.

\subsection{Tunneling of 5 min oscillations to the chromosphere}
\vspace{0.2cm}

A specific question for the models of wave propagation in small scale
structures is the explanation of the power spectrum of waves at chromospheric
heights (see Section~\ref{sect:obs}).
Observations in network regions mostly agree that long-period 5 min
oscillations are transmitted to the chromosphere in the close proximity of
network magnetic elements (Section~\ref{sect:network}). In a plage, it seems
not clear yet whether these long-period oscillations are transmitted at the
plage borders with predominantly inclined field or vertically from the
photosphere (Section~\ref{sect:plage}).

There have been a number of works arguing that inclined magnetic field at the
boundaries of enhanced network and plage regions allows long-period slow
acoustic waves to propagate into the solar chromosphere due to a reduction of
the cut-off frequency \cite{DePontieu+etal2004, DePontieu2005, Hansteen2006,
Jefferies+etal2006, McIntosh+Jefferies2006}. Numerical 1D modeling confirms
the efficiency of this mechanism \cite{Heggland2007}: long-period 5 min waves
can only propagate upwards where the magnetic field is inclined, whereas
waves with periods around 3 min dominate when the field is vertical. This
modeling includes a complex treatment of radiative transfer, but
simplifications of a one-dimensional propagation in a constant inclined
field.

Alternatively, some authors proposed that radiative losses in thin flux tubes
can help propagating long-period oscillations vertically upward
\cite{Centeno+etal2009, Khomenko+etal2008b}. The presence of radiative losses
can lead to a significant reduction of the cutoff frequency, formally adding
a propagating component to the oscillations \cite{Roberts1983}.
Two-dimensional simulations of flux tube waves with a 5 min periodic
photospheric driver, and including radiative losses by means of a simplified
Newton's law with fixed radiative relaxation time, demonstrate the
effectiveness of this mechanism in leaking 5 min oscillations into the
chromosphere inside vertical flux tubes \cite{Khomenko+etal2008b}. The power
and phase spectrum of waves in these simulations reproduce observations in
facular regions \cite{Centeno+etal2009}.

The treatment of radiative losses by Newton's law \cite{Khomenko+etal2008b}
is very rough. Recently, realistic radiative magneto-convection simulations
were performed in a model extending from the convection zone to the corona,
where oscillations were produced self-consistently by the turbulent motion in
the convection zone without any imposed external driver \cite{Heggland2011}.
A sophisticated state-of-the-art treatment of radiative losses was included.
The simulations aimed at quantifying the contribution of both effects - field
inclination and radiative losses - into the resulting spectrum of waves at
chromospheric heights. It was found that 5 min oscillations are able to
propagate in regions where the field is inclined and strong, including the
edges of otherwise vertical flux tubes, whereas in regions where the field is
vertical (center of flux tubes) or weak, oscillations with periods around 3
min propagate \cite{Heggland2011}. Radiative losses were found to play only a
minor role in determining the propagation spectrum (though no reference
simulation with radiative transfer switched off was performed).

There are also alternative mechanisms for transmission of 5 min oscillations
into the chromosphere, by means of non-linear photospheric pulses that create
nonlinear wakes leading to a trail of consecutive shocks with 5 min
periodicity \cite{Zaqarashvili2011b}.

\subsection{Power distribution at different frequencies}
\vspace{0.2cm}

Many recent studies agree that the peculiar power shadowing and enhancements
observed in plage and network areas are related to processes of
mode conversion and refraction at the inclined magnetic canopy in the
proximity of the $\beta=1$ layer (see Sections~\ref{sect:network} and
\ref{sect:plage}).

There have been numerous theories to explain high-frequency acoustic halos in
plage regions \cite{Brown+etal1992, Braun+etal1992, Hindman+Brown1998,
Jain+Haber2002, Kuridze2008, Jacoutot+etal2008, Hanasoge2008,
Khomenko+Collados2009}. Numerical simulations of interaction of waves with
sunspot/plage flux tubes suggest that the power enhancement in halos is
produced by mode mixing induced by the magnetic field, resulting in
preferential scattering from low to high wave numbers \cite{Hanasoge2009}.
Alternatively, an explanation has been offered based on the enhanced
high-frequency wave excitation at the regions of intermediate magnetic field
strength in halos \cite{Jacoutot+etal2008} .

Single-source numerical simulations of magneto-acoustic wave propagation in a
magneto-static sunspot model demonstrate that the halo effect happens in a
natural way due to the additional energy input from the high-frequency fast
mode waves, after their refraction above the $\beta=1$ layer
\cite{Khomenko+Collados2009}. The frequency and location of the halo is
extremely sensitive to the relative location of the $\beta=1$ layer, the
cut-off layer and the formation height of a given observed spectral line. The
halo is produced in those photospheric regions where the field is
intermediate, implying that the Alfv\'en speed is lower than the sound speed.
The halo is not observed at low frequencies because these waves are already
reflected by cut-off below the transformation layer, and is not observed in
the umbra of the sunspot because the refraction happens below the layer
visible to spectral line observations. This model
\cite{Khomenko+Collados2009}, as well as the model of mode mixing
\cite{Hanasoge2009} suggest that the halo power enhancements might be larger
in the horizontal velocity component.


To clarify the nature of magnetic shadows, numerical simulations of
high-frequency magneto-acoustic wave propagation in a snapshot of a
three-dimensional model of magneto-convection were carried out
\cite{Nutto2012}. On their way up, acoustic waves were converted into
different mode types and were refracted. The Fourier analysis of this
simulation clearly shown chromospheric shadows, similar to the observed ones.
It was concluded that these shadows are linked to the mode conversion process
and that power maps at these heights show the signature of the different
magneto-acoustic wave modes. Yet another mechanism to explain power
enhancements in halos was proposed based on acoustic waves trapped in
field-free atmospheres lying below small-scale magnetic canopies of network
cores and active regions \cite{Kuridze2008}.

Alternative pictures of wave transformation and refraction \cite{Kuridze2008,
Khomenko+Collados2009, Nutto2012} are, in fact, similar, suggestive that
both, halos and shadows, are different manifestation of the same physical
mechanism. Recent observational results seem to favor the model based on wave
transformation and refraction as well \cite{Rajaguru2012}.
One can not exclude, though, that other effects may also play a role
\cite{Hanasoge2009, Jacoutot+etal2008}. For example, acoustic halos are found
to be co-spatial with acoustic glories, locations with enhanced seismic
emission surrounding active regions \cite{Donea2011}. The nature of the
glories is different to halos, since they tell us about an increase of
seismic emission, i.e. power of waves emanating directly from the wave
sources, whereas the acoustic halos are just measures of increased wave
amplitudes, whatever is the reason. Thus, the last word on the nature of
halos is not yet said.

\section{Conclusions}
\vspace{0.3cm}

Significant advances have been made in our understanding of the physics of
atmospheric waves and their interaction with local magnetic structures in the
photosphere, chromosphere and above. Summarizing the vast amount of
theoretical and observational works reviewed here, we give below some brief
conclusions and suggestions for perspective future work directions:

\begin{itemize}

\item The observed distributions of photospheric and chromospheric power
    of long and short period waves seems to be distinct over network and
    plage/facular regions. In the network, long-period 5 min waves seem
    to be transmitted in the close proximity of magnetic elements, while
    short-period 3 min waves are ``shadowed'' due to the interaction of
    magneto-acoustic waves with the more horizontal fields of the
    canopies. In the stronger-field plage regions, short-period (3 min,
    $\nu=5-7$ mHz) halos dominate both in the photosphere and in the
    chromosphere, and the propagation of long-period waves is enhanced
    for inclined fields, but vertical propagation has also been reported.

    The reason for that different behavior may be the variation of the
    height of the magnetic canopy $\beta=1$ layer, being lower for
    stronger plage fields, thus modifying the spectrum of waves reaching
    the heights sampled by photospheric and chromospheric observations.

    More high-resolution studies are needed in this direction, comparing
    regions with different magnetic fluxes, both at the disc center and
    closer to the limb to get information about horizontal velocities.
    Simultaneous measurements of the magnetic field vector are also very
    important.

\item On the theoretical side, there seems to be still no agreement on
    the explanation of the presence of ubiquitous long-period waves in
    the chromosphere above network bright points and plage. Different
    observations give evidences both for inclined and vertical
    propagation (both upward and downward), at least for the network
    elements, while theoretical models seem to point toward the inclined
    propagation as a dominant mechanism. Inclined propagation receives
    more observational support especially for the strong field plage
    regions where it is expected that the slow acoustic waves propagate
    field-aligned in the low-$\beta$ regime. Future studies should
    address the propagation in weaker structures, accompanied by
    observations of more network and inter-network elements.

\item There have been few detections of magnetic field oscillations in
    quiet solar regions reported in the recent literature. Magnetic field
    and flux oscillations are very hard to detect, but an effort should
    be made to increase the number of such studies, since the amplitude
    and phase relations between the magnetic field variations and other
    parameters give valuable information on the observed wave types.

\item Idealized simulations of wave propagation from the photosphere to
    the chromosphere in small scale flux tubes are rather well developed.
    These models have suggested a number of mechanisms to transport the
    energy of the wave to the upper layers. Three-dimensional simulations
    are being developed. A remaining challenge is to couple the
    sub-photospheric layers, photosphere, chromosphere, transition region
    and corona by means of waves. This is needed to understand how (and
    what amount of) the wave energy can reach the corona, depending on
    the driver properties, and taking into account all the wave physics
    such as 3D mode transformation, refraction, reflection, non-linear
    interaction, as well as dissipation mechanisms. It will be
    interesting in the future to get more insights on the efficiency of
    conversion to Alfv\'en waves on inclined canopy fields of the quiet
    Sun small scale flux tubes, by means of 3D numerical simulations.

\item Oscillations in the surroundings of quiet Sun magnetic structures
    in network, plage and facular regions are being observed at extremely
    high spatial resolution, and their relation to the magnetic topology
    is being clarified. Theoretical models for acoustic halos and shadows
    indicate that both may have a similar origin, tracing mode
    transformation process, and fast mode refraction and reflection at
    the inclined field of the magnetic canopy. It remains to clarify the
    relation between the acoustic halos and glories, present on the same
    locations. Some theoretical models for halos suggest that the power
    enhancement in horizontal velocity component should be significantly
    stronger than in the vertical one. In the future it will be
    interesting to study halos and shadows at locations off the disc
    center to shed more light on their nature. Models of transformation
    to Alfv\'en waves suggest they will be reinforced at the periphery of
    active regions. Checking this in future observations is important as
    well.

\end{itemize}

\vspace{0.3cm}

\hspace{-0.5cm}{\bf Acknowledgements:} This research has been supported by
the Spanish Ministry of Economy and Competitiveness (MINECO) under the grants
AYA2010-18029 and AYA2011-24808.

\providecommand{\newblock}{}


\end{document}